\documentclass[usenatbib]{mn2e}
\usepackage{graphicx}
\usepackage{rotating}

\newcommand{\lsol}{\hbox{$L_\odot$}}

\newcommand{\e}[1]{$10^{#1}$}
\newcommand{\ee}[1]{$\times 10^{#1}$}         
\newcommand{\kms}{~km\,s$^{-1}$} 
\newcommand{\cm}[1]{~cm$^{#1}$}
\newcommand{\ergs}{~erg\,s$^{-1}$\,cm$^{-2}$\,sr$^{-1}$}
\newcommand{\erg}{~erg\,s$^{-1}$\,cm$^{-2}$}

\newcommand{\co}{$^{12}$CO}    
\newcommand{\oh}{OH(1720~MHz)~}
\newcommand{\tco}{$^{13}$CO}
\newcommand{\htco}{H$_2$CO}
\newcommand{\hcop}{HCO$^+$}

\newcommand{\hone}{2.12~$\micron$ H$_2$ 1--0 S(1)\ }

\newcommand{\h}{H$_2$}
\newcommand{\hi}{H\,{\sc i}}

\newcommand{\tmb}{$T_{mb}$}

\newcommand{\mb}{main-beam brightness temperature\ }

\newcommand{\gsim}{\,\raisebox{-0.4ex}{$\stackrel{>}{\scriptstyle\sim}$}\,}
\def\micron{\hbox{$\mu$m}}

\newcommand{\aap}{A\&A}

\newcommand{\aj}{AJ}
\newcommand{\apj}{ApJ}
\newcommand{\apjl}{ApJL}
\newcommand{\apjs}{ApJS}
\newcommand{\mnras}{MNRAS}

\newcommand{\nat}{Nature}

\begin{document}

\title[Shocked molecular hydrogen towards the Tornado nebula]
{Shocked molecular hydrogen towards the Tornado nebula}

\author[Lazendic et\ al.]
       {J. S. Lazendic$^{1,2}$,
       M. Wardle$^3$, M. G. Burton$^4$, F. Yusef-Zadeh$^{5}$, \newauthor 
 A. J. Green$^1$ and J. B. Whiteoak$^6$  \\
$^1$ School of Physics A28, University of Sydney, Sydney NSW 2006, Australia\\
$^2$ Harvard-Smithsonian Center for Astrophysics, 60 Garden Street, Cambridge, MA 02138, USA\\
$^3$ Department of Physics, Macquarie University, N SW 2109, Australia\\
$^4$ School of Physics, University of New South Wales, Sydney NSW 2052, Australia\\
$^5$ Department of Physics and Astronomy, Northwestern University,
       Evanston, IL 60208, USA\\
$^6$ Australia Telescope National Facility, CSIRO, PO Box 76, Epping
       NSW 1710, Australia }

\date{ }


\maketitle


\begin{abstract}

We present near-infrared and millimetre-line  observations of the
Tornado nebula (G357.7--0.1).  We detected \hone  line emission
towards the suspected site of interaction with a molecular cloud
revealed by the presence of an \oh maser.  
The distribution of the \h\ emission is well correlated with the
nonthermal  radio continuum  emission from the Tornado, and the
 velocity of the \h\ emission spans over 100\kms, which both 
imply that the \h\
emission  is shock excited. We also detected millimetre-lines  from
\co\ and \tco\ transitions at the velocity of the maser,  and mapped
the distribution of the molecular cloud in a  $2\times2$~arcmin$^2$
region around the maser. The peak of the  molecular cloud aligns
well with an indentation in  the nebula's  radio continuum distribution,
suggesting that the nebula's shock  is being decelerated at this
location, which is consistent with the presence of the \oh maser and
shocked \h\ emission at that location.

\end{abstract}


\section{Introduction}

The inner region of our Galaxy is rich in unusual and unique
sources. One such object is the Tornado nebula (G357.7--0.1,
MSH~17--39), a peculiar nonthermal source with an axially symmetric
morphology. Because of its filamentary structure, steep nonthermal
radio spectrum and presence of significant linear  polarisation, the
Tornado has been classified as a supernova remnant (SNR)
\citep{milne79,clark76,shaver85-1,stewart94}. Its elongated and loop-like
structure was suggested to be a result of biconical flows from  the
progenitor star  \citep{manchester87} or an accreting binary system
\citep{helfand85,becker85}, but no associated pulsar or neutron star
has been found.  X-ray emission has been detected from the
nebula with the {\em Advanced Satellite for Cosmology and
Astrophysics (ASCA)} \citep{yusef03} and 
{\em Chandra X-ray Observatory} \citep{gaensler03}. 
Although the emission from the nebula is most probably thermal,
deeper observations are needed to unambiguously establish the nature of
X-ray emission.

A compact source, called the `Eye' of the Tornado, located about  30
arcsec west from the Tornado, has been linked to the nebula and
suggested to be a high proper-motion pulsar associated with the nebula
\citep{shull89}. However, the  source was found to have a flat radio
spectrum suggestive of an H\,{\sc ii} region \citep{shaver85-2}. Most
recently, \citet{burton03} found 2.16 \micron\ Br$\gamma$ emission
towards the Eye which peaks at a velocity of around $-200$\kms. These
observations imply that the Eye is an isolated core where an embedded
massive star is in the process of formation, and it is  a foreground
object to the Tornado. Hydrogen recombination lines were also 
detected towards the Eye at radio wavelengths \citep{brogan03}.

\oh masers have been found at the northwestern tip of the Tornado
\citep{frail96,yz99}.  When not accompanied by detectable maser
emission from the other three OH ground-state transitions at 1612,
1665 and 1667~MHz,  the 1720~MHz OH transition is believed to be an
indicator of an interaction between SNRs and molecular clouds
\citep*[e.g.][]{frail96,green97,lock99,wardle99,wardle02}. The maser in the
Tornado has a velocity of $-12.4$\kms, and if its distance is assumed
to be more than a few hundred parsec  (avoiding the distance
ambiguity), a distance of 11.8~kpc  is derived from the Galactic
rotation curve \citep{fich89}.  This is consistent with  21-cm \hi\ 
absorption against the Tornado, which places the nebula
at a distance greater than 5~kpc \citep{Radhak72}.  The magnetic field
strength towards the \oh maser, determined from measurements of
OH-line Zeeman splitting, is 0.7~mG \citep{brogan00}.

We obtained near-infrared (NIR) and millimetre-line observations towards
 the northwestern part of the Tornado to investigate the implied  interaction
 of the nebula and the surrounding molecular gas. We also obtained
 archival radio continuum data at 20~cm for a comparison. The observations 
are described in Section~2 and the results are given in
 Section~3. The results are discussed in Section~4 and 
summarized in Section~5.

\section{Data}
\label{sec-t2}

\subsection{UNSWIRF observations}
\label{sec-unswirf}

For NIR observations of the Tornado we used the University of New
South Wales Fabry-Perot narrow-band tunable filter
\citep[UNSWIRF;][]{ryder98} mounted on the  3.9-m Anglo-Australian
Telescope during June 1999. The observations were centred near the
location of the \oh maser in the nebula at  ${\rm RA}(1950)=17^{\rm
h}~36^{\rm m}~55\fs 4$, ${\rm Dec.}(1950)=-30\degr~56\arcmin~1\farcs
5$. Only a small part of the Tornado was covered because the UNSWIRF
aperture had a circular image of 100 arcsec in diameter with a pixel
size of 0.77 arcsec. The UNSWIRF field of view is marked with a circle
in Figure~\ref{fig-radio}. Five frames at different Fabry-Perot
settings were obtained in the \hone line, equally spaced by 40\kms\
and centred on the maser velocity. The velocity resolution of the
instrument was $\approx 75$\kms. The integration time was 180 seconds
per frame.  For the continuum subtraction, an additional frame was
taken at a velocity setting of $-400$\kms\ from the first frame.  A
velocity cube, constructed from the five frames, is fitted with  the
instrumental Lorentzian profile to determine  the \h\ line parameters
across the field.  The \h\ line flux density and line centre velocity
 are determined to within $\approx 30$ per cent and $\approx
20$\kms, respectively. The intensity was calibrated using additional
observations of the standard star BS~8658. To establish the coordinate
scale for the UNSWIRF image, we used the $K$-band (2.15~$\micron$)
stellar positions listed in the Two Micron All Sky Survey (2MASS)
point source catalogue\footnote{see
http://spider.ipac.caltech.edu/staff/hlm/2mass/calvsuc/calvsuc.html}
 \citep{cutri97},  which are accurate to 0.1~arcsec.

\subsection{SEST observations}

We used the 15-m Swedish-ESO Submillimeter Telescope (SEST) during
June 2000 to observe the region towards the \oh maser in  \co, \tco,
CS, \hcop, HCN and \htco\ transitions between 1.3 and 3~mm.
Table~\ref{tab-torn-sest} lists the observed molecular transitions,
frequencies and corresponding telescope beam widths.  The spectral
resolution ranged between $\approx 0.1$\kms\ at 1.3~mm and  $\approx
0.3$\kms\ at 3~mm.  The observations were performed in
position-switching mode with reference position at ${\rm
RA}(1950)=17^{\rm h}~47^{\rm m}~10^{\rm s}$, ${\rm
Dec.}(1950)=-32\degr~43\arcmin~00\arcsec$.  The two \co\ transitions
were obtained over a  $2\times2$ arcmin$^2$ region centred at  ${\rm
RA}(1950)=17^{\rm h}~36^{\rm m}~54^{\rm s}$, ${\rm
Dec.}(1950)=-30\degr~56\arcmin~25\arcsec$, with a 20 arcsec grid and
30 second integration per position.  For the other molecular
transitions, observations were taken in a five point-cross grid
centred at  ${\rm RA}(1950)=17^{\rm h}~36^{\rm m}~54^{\rm s}$, ${\rm
Dec.}(1950)=-30\degr~56\arcmin~05\arcsec$, plus 20 arcsec offset.  The
integration times were 60 seconds for the \co\ and \tco\ transitions
and 180 seconds for all the other transitions. 

 Telescope pointing accuracy to better than 5 arcsec was maintained
using periodic observations of the SiO masers associated with  AH\,Sco
and W\,Hyd. The data have been corrected on-line for atmospheric
absorption by periodic observations of a blackbody calibration
source. After baseline subtraction and Hanning smoothing, the data
were  scaled to main-beam brightness temperature (\tmb) using the main-beam
efficiencies (0.74, 0.70, 0.67 and 0.45 at 85--100~GHz, 100--115~GHz, 
130--150~GHz and 220--265~GHz respectively).

\subsection{VLA data}

The radio continuum data were obtained from archives of the Very Large
Array (VLA) of the National Radio Astronomy Observatory (NRAO). The
observations  of the Tornado nebula were carried out on 28 April 1991
and  13 December 1985 at 20~cm in its A and D-array  configurations,
respectively. Standard calibration of both data sets used 1748--253
and 3C286 as the phase and amplitude calibrators,
respectively. Standard self-calibration was also applied to both data
sets before  the final  image was constructed using the Maximum Entropy
method (MEM in {\sc aips}).

\section{Results and analysis}
\label{sec-t3}

The 20~cm radio image of Tornado, shown in Figure~\ref{fig-radio}, has
a resolution of $2.65\arcsec  \times 1.07\arcsec$ (PA=$-6.7\degr$).
The image illustrates nebula's filamentary structure and we indicate
the main features --  the bright western region is  called the
`Head', which is followed by  a prominent arc in the centre of the
nebula and faint filaments comprising the `Tail' of the nebula. The
compact source near the nebula, the Eye of the Tornado, is also
indicated.

\subsection{\h\ emission}

We have detected for the first time \h\ line emission towards the
Tornado. Figure~\ref{fig-h2} shows contours of the
velocity-integrated \hone line emission overlaid on a greyscale image
of the 2MASS $K$-band ($2.15\micron$). The emission is ring-shaped
with a diameter of $\approx 50$ arcsec, and contains a number of
peaks. The emission
peaks are 5--10 arcsec in size. The peak flux density of the \hone\
line emission is 1.4\ee{-4}\ergs\ and the 3$\sigma$ rms noise is
8\ee{-5}\ergs; the integrated flux density of
the whole source is 2.4\ee{-12}\erg. After correcting for an assumed
extinction in $K$-band of at least $A_K\approx3$ (since the object is
behind the Galactic centre), the peak flux density of the \hone\ line
emission is 2.2\ee{-3}\ergs, the integrated flux density is
3.8\ee{-11}\erg, and the 2.12$\micron$ \h\ luminosity is 167\lsol\ for
distance of 11.8~kpc.

The \h\ line centre velocities, marked in Figure~\ref{fig-h2} span
 over $\sim 100$\kms\ across the emission region. There are 7
 identifiable clumps, with velocities ranging from $-95$ to +25\kms,
 with the mean velocity around  $-54$\kms. They range from 40\kms\
 blue of the mean to 70\kms\ red of the mean. Apart from the clump at
 +26\kms, however, they are all blue-shifted from the velocity of the
 cloud, as determined by the CO and OH maser emission. The emission at positive velocities appears as a knot located at
the south-eastern region. Unfortunately, this component was not observed with
sufficient Fabry-Perot spacings to infer the red-shifted
extent of its emission.

\subsection{CO emission}

Several \co\ and \tco\ emission components at different velocities
were detected along the line of sight to the Tornado nebula. 
 Some of the features can be seen in Figure~\ref{fig-spectra}, which 
 shows the \co\ 1--0 spectrum  towards the position of the 
\oh maser in the range $-40$ to +15\kms.  
In Figure~\ref{fig-spectra-co} we show the individual \co\ 2--1
spectra across a $2 \times 2$~arcmin$^2$ region of the Tornado's
Head. The  (0,+20) position corresponds to the location of the \oh maser.
Similarly, Figure~\ref{fig-spectra-tco} shows the five \tco\ 2--1
spectra, where the (0,0) position corresponds to the location of the \oh
maser.
We have assumed that the molecular cloud associated with the Tornado
nebula has a velocity close to  the maser 
velocity ($-12.4$\kms), and the spectrum shows a CO feature 
near $-12$\kms\ with a linewidth of $\sim 4$\kms. 
The spectral features were fitted with Gaussian profiles
and the resulting line parameters for the cloud associated with the
nebula  are summarised in Table~\ref{tab-torn-sest}. In contrast  
to the \co\ and \tco\ transitions, the other
molecular transitions were found to be very weak, 
providing only upper limits (Table~\ref{tab-torn-sest}). 
We found no significant line broadening ($\gsim 10$\kms) 
in the present data which
would provide a clear kinematic evidence of shock interaction.

The measured line ratio 
$R=T^{^{12}{\rm CO}}_{mb}/T^{^{13}{\rm CO}}_{mb}$ was 
$\approx 3.4$ for the 1--0 transitions, and $\approx 3.0$ for the 2--1
transitions, which implies
that the \co\ emission is optically thick. Adopting an overall isotope
ratio $R_I=[^{12}{\rm C}/^{13}{\rm C}]= 50\pm20$
\citep[e.g.,][]{langer90}, we estimate an upper limit on the \co\ optical
depth of $\tau \approx R_I/R\approx 21$ 
for the 1--0 line. Since \co\ is optically thick, the actual 
\co\ 1--0/2--1 line ratio should
be close to unity. Thus, by equating the two brightness
temperature of the two \co\ lines, we derived a
beam-deconvolved source size of $\approx 40$~arcsec. We can then
calculate the actual brightness temperatures corrected for the  source
size from $T_S=T_{mb}(1+\Omega_B/\Omega_S)$, where $\Omega_B$ and
$\Omega _S$ are the beam and source solid angles, respectively
\citep*[e.g.,][]{jan94,rohlf96}. For the two \co\ transitions we
 then have $T_S(1-0)=20$~K and $T_S(2-1)=14$~K. To determine the
temperature of the molecular gas towards the \co\ peak, we used the
molecular-line excitation code with the mean escape probability (MEP)
\citep[for details see e.g.,][]{jan94}. Our excitation modeling of the two
\co\ lines  implies a kinetic temperature of $20\pm5$~K for a gas
density of \e{4}--\e{6}\cm{-3}. This is consistent  with the fact
that for an optically thick \co\ the gas temperature should be similar
to the brightness  temperature of the \co\ lower-energy transitions.  
We also estimate \co\ column density
of $N$(\co)$\approx (1-8)\times10^{17}$\cm{-2} from our modeling.  Using 
the standard fractional
abundance [\co]/[\h] of \e{-4}\  \citep*[e.g.,][]{irvine87,vand93} we
derive $N$(H$_2) \approx (1-8)\times 10^{21}$\cm{-2}.  The upper limit
of the \tco\ optical depth is $24/70 = 0.3$, and column
density is 2.7\ee{16}\cm {-2}. Because of the non-detection of molecules
with larger dipole moments (e.g., CS), we do not have a
 more reliable method to constrain the density of the molecular gas.

A velocity-integrated map of \co\ 2--1 emission 
between $-10$ and $-16$\kms\ is shown in 
Figure~\ref{fig-co-map}.  Since the optical
 depths are large, the observed \co\ distribution traces the
 temperature distribution rather than the column density distribution.

\section{Discussion}
\label{sec-t4}

\h\ emission has been studied in only a few of the known SNR
masers, such as IC~443 \citep{burton88}, Sgr~A East \citep{fyz01} and
G359.1--0.5 \citep{lazendic02}.
The  line ratio of the \h\ 1--0 and 2--1~S(1) transitions may be used
to establish the excitation mechanism for \h\ emission.
Unfortunately, we were unable to obtain 2--1~S(1) observations towards
the Tornado, but we argue on the basis of morphology and kinematics
that the \h\ emission detected towards the \oh maser in the Tornado
nebula is shock excited, as found in other SNR associated with \oh
emission \citep[e.g.,][]{burton88,lazendic02,lazendic03}.
In Figure~\ref{fig-h2+radio} we compare the integrated 
\hone line emission with the 20~cm radio continuum  emission. In the 
region where the SNR encounters a dense molecular cloud the 
 SNR shock will be decelerated, producing enhanced radio 
emission due to strong shock compression and enhanced particle acceleration. 
 Both the \h\ and radio continuum emission features 
have filamentary morphology, and there is a good
correlation between the \h\ and radio continuum maxima; we
interpret this agreement as  an indication of physical
association. The bright clumps in the \h\ emission represent
small-scale structures in the inhomogeneous ambient medium,
illuminated by the SNR shock \citep[see e.g.,][]{burton88}.
Another argument for shock excitation is the significant velocity
motion in the \h\ emitting gas across the region, which is opposite to
that found in fluorescent sources where velocities are essentially at
the ambient velocity of the molecular cloud \citep{burton90}.  The \h\
emission is also spatially coincident with extended \oh emission found
towards the nebula \citep{yz99}, which is believed to be evidence of
large-scale interactions between  a SNR and an adjacent molecular
cloud.

The molecular cloud associated with the nebula and the \oh maser is
optically thick, with a kinetic temperature of $\approx 20$~K,
assuming a gas density of \e{4}--\e{6}\cm{-3}. This 
temperature and density are in
agreement with properties expected for the molecular cloud interacting
with SNRs associated with \oh masers \citep{lock99}, which suggests
that we are detecting mostly the pre-shock molecular gas.  
We did not detect the broad line profiles from
the warm post-shock region, like those seen in IC~443 \citep{vand93}. 
The 50~arcsec diameter of the \h\ ring is comparable
to the 45~arcsec beam size at the \co\ 1--0 frequency, and we might
expect to detect the warm shocked gas. Indeed, the location of the \oh
maser within the CO peak contour in Figure~\ref{fig-co-map} 
indicates that we might detect  contribution from both the pre-shock and
post-shock gas. However, the molecular gas
is optically thick and we might be probing only the cool envelope of the
cloud with the observations of the \co\ transitions. Furthermore, 
weaker broad wings from the warm gas might be confused with neighbouring molecular
features. Similar observations of shocked molecular gas towards
other SNRs imply that the broad emission component is usually stronger
in higher \co\ transitions \citep[e.g.,][]{seta98}, or restricted to a
small molecular clumps \citep[e.g.,][]{reach02}. Thus,   observations
of higher energy transitions  (which would also result in higher
spatial resolution) might be needed to detected the shocked molecular
gas in the millimetre band.  Indeed, preliminary results from
observations of CO 4--3 line show broad line emission centered at
$-12$\kms\ (Yusef-Zadeh et al. 2004, in preparation).

Figure~\ref{fig-h2+co}a shows the integrated \co\ 2--1 line
emission superimposed on the 20~cm radio continuum  emission.  The
blast wave delineated by the radio emission is indented at the
location of the \oh maser and the \co\ peak, suggesting that the
molecular cloud is large and dense enough to cause a significant impediment to
the shock front. This resembles the well known case of SNR-cloud
interaction seen in the Cygnus Loop \citep[e.g.,][]{danforth01}. Also,
the western side of the \h\ emission appears more flattened compared to
the other parts, which is suggestive of a compression
resulting from the interaction between the nebula and the surrounding
molecular gas.  Thus,  the \oh maser, the \h\ emission and the radio
continuum indentation is the indication that the blast wave has just
begun to interact with the tip of the molecular cloud. The
distribution of molecular gas components is consistent with the \oh
maser being located in a post-shock layer just behind a shock front,
oriented transverse to the line of sight, as predicted by the theory
for \oh maser formation \citep{elitzur76,lock99}.  
Furthermore, the X-ray
flux observed from the Tornado \citep{gaensler03} 
is consistent with that required to produce the necessary OH \citep{wardle99}.

The \oh maser is located at the western edge of the \h\ emission and its
velocity ($-12$\kms) differs significantly from the mean velocity of
\h\ emission. This is not too surprising because we see masers only at
the location where the shock is propagating perpendicular to our line
of sight \citep{lock99}, whereas we see \h\ emission from any
location in the shocked gas. The extended maser emission spans over
a larger velocity gradient between $-13.1$ and $-8.7$\kms\ \citep{yz99}. It is
expected that the shock associated with extended maser emission is not
exactly perpendicular to the line of sight, although these velocities
are still much lower than the mean velocity of \h\ emission.
The large velocity span of \h\
emission features is probably the result of {seeing a shock under
different angles. Although \h\ emission appears ring-like, the
distribution of \h\ line-velocities, when compared to $-$12\kms\ as a 
reference, is clearly incompatible with any expansion of a ring.  
In fact, it may be simply
a chance line of sight projection that is causing the apparent ring
structure. Presumably the north-west structure of the \h\ ring 
is marking the shock front as it runs into the CO cloud, given the 
position offset between the CO and \h\ in Figure~\ref{fig-h2+co}b.  
The rest of the \h\ ring is probably not a continuous structure but 
  foreground or background structure being shocked with a 
different shock velocity vector. This is supported by the fact that
there is no coherence between two adjacent 
clumps at different velocities when we examine the \h\ cube, and
velocities  jump from one clump to the other, supporting the notion
that clumps are distinct features. 
If we suppose the greatest shock speed is $V_{max}$, running in to the 
lowest density molecular gas, $n_{min}$, then the line of sight velocity we 
measure is $V = V_{max} (n_{min}/n)^{0.5} cos(\theta)$, where 
the velocity $V$ is with respect to the $-12$\kms\ ambient cloud value,
 with $\theta = 90\degr$ being the plane of the sky, and $n$ is 
the density at that point. In 
general the velocities get more negative towards the rear of the ring, 
which implies that $\theta$ is small here so the shock is moving nearly 
directly towards us to the rear of the \h\ ring.

\section{Summary}
\label{sec-t5}

We have detected  \hone emission towards the northwestern edge of the
Tornado nebula.  The emission has a ring-like morphology and
encompasses the location of the \oh maser. The correlation between
the radio continuum  and \h\ emission, and significant velocity
motions  suggest that the \h\ emission originates from an expansion of
the shock wave and is most probably shock excited. The \oh maser is
located at the western edge of the \h\ emission, which probably 
delineates the leading edge of shock front.

Molecular transitions of \co\ and \tco\ were also detected at the
maser velocity of $-12$\kms; emission from CS, \hcop, HCN and \htco\ 
was not detected. The millimetre-line
observations were found to probe only the ambient molecular gas
associated with the nebula, which is optically thick and warm ($\approx
20$~K, for assumed gas density of \e{4}--\e{6}\cm{-3}). The 
location of the molecular
gas coincides with an indentation in the radio continuum distribution
and the \oh maser position, suggesting that the molecular cloud is
large enough to cause significant deceleration of the blast wave
shock.  Observations of optically thinner higher transitions of \co\
are needed to examine the properties of shocked part of the molecular
gas.

\section{Acknowledgments}

We thank John Black for kindly providing us with his MEP code.
JSL was supported by an 
Australian Government International Postgraduate Research Scholarship,
 a Sydney University Postgraduate  Scholarship, and an ATNF
Postgraduate Scholarship. JSL also acknowledges 
travel support from the Australian Government's 
Access to Major Research Facilities Program.

The NIR observations would not have been 
possible without the efforts of Michael
Ashley and the UNSWIRF crew from UNSW, as well as the staff of 
the Anglo Australian Observatory. 
The Swedish-ESO Submillimetre Telescope (SEST) is operated by the
Swedish National Facility for Radio Astronomy, Onsala Space
Observatory and by the European Southern Observatory (ESO).
The National Radio Astronomy
Observatory (NRAO) is a facility of the National Science Foundation, operated
under a cooperative agreement by Associated Universities, Inc.
This publication makes use of data products from the Two Micron All Sky Survey
(2MASS), which is a joint project of the University of Massachusetts
and the  Infrared Processing and
Analysis Center/California Institute of Technology, funded by the
National Aeronautics and Space Administration and the National Science
Foundation.

\clearpage

\begin{table*}
\centering
\caption{Molecular lines observed towards the Tornado nebula 
at  ${\rm RA}(1950)=17^{\rm h}~36^{\rm m}~54^{\rm s}$, ${\rm
Dec.}(1950)=-30\degr~56\arcmin~05\arcsec$. The 
first three columns list the molecular
transitions and their frequencies. The fourth column lists the SEST beam sizes (FWHM),  and the last three columns give the spectral-line parameters: \mb (\tmb) and rms noise (upper limits are 2$\sigma$ values), central velocity ($V_{LSR}$) and line width ($\Delta V_{LSR}$).}
\begin{tabular}{@{}lcccccc}
\hline\hline
Molecule & Transition & $\nu$ & Beam Size & $T_{mb}$ & $V_{LSR}$ & 
$\Delta V_{LSR}$  \\
 &  & (GHz) & (arcsec) &  (K)  & (\kms) & (\kms) \\
\hline
\co   & 2--1 & 230.538 & 23 & 10.5$\pm$0.3 & $-$11.0 & 4.0  \\ 
      & 1--0 & 115.271 & 45 &  8.6$\pm$0.4 & $-$11.2 & 5.1 \\
\tco  & 2--1 & 220.399 & 23 &  3.5$\pm$0.2 & $-$11.0 & 3.7\\ 
      & 1--0 & 110.201 & 45 &  2.5$\pm$0.1 & $-$11.1 & 3.7 \\
CS    & 3--2 & 146.969 & 34 &  $<$0.1 & -- & -- \\
      & 2--1 & 97.981  & 52 &  $<$0.2 & -- & --  \\
\hcop & 1--0 & 89.188  & 54 &  $<$0.2 & -- & --  \\
HCN   & 1--0 & 88.632  & 55 &  $<$0.1 & -- & -- \\
\htco & 3$_{(2,2)}$--2$_{(2,1)}$ & 218.475 & 23  & $<$0.1 & -- & -- \\ 
      & 3$_{(0,3)}$--2$_{(0,2)}$ & 218.222 & 24 &  $<$0.1 & -- & -- \\
     
\hline
\end{tabular}
\label{tab-torn-sest}
\end{table*}


\clearpage

\begin{figure*}
\centering
\includegraphics[height=10cm]{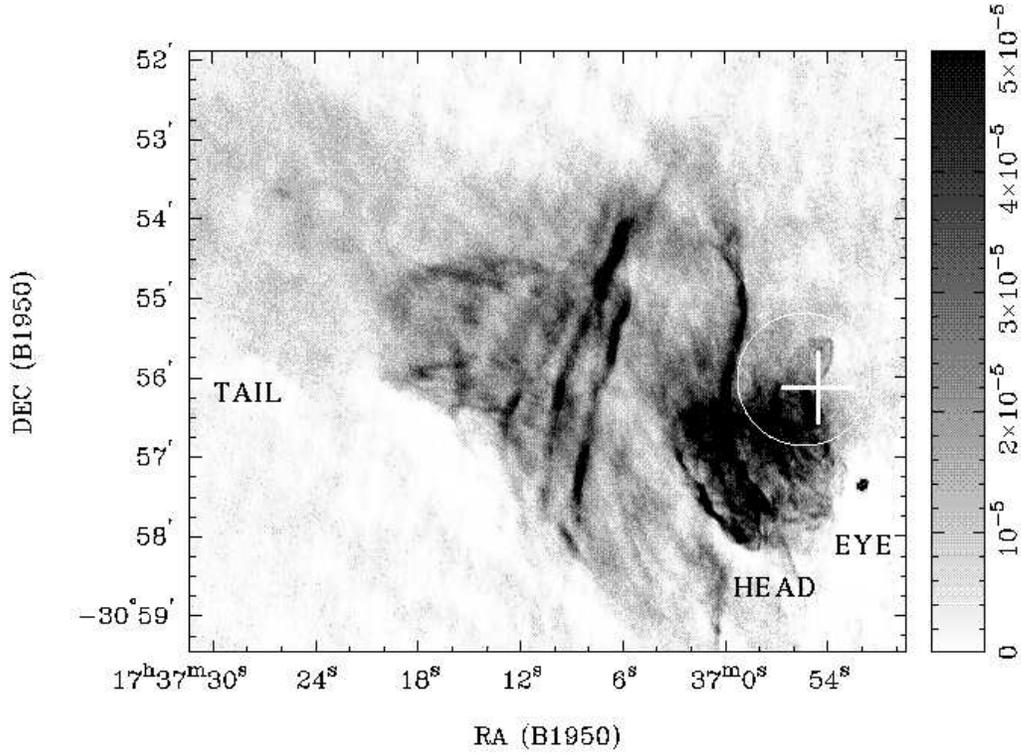}
\caption{20~cm Very Large Array image of the Tornado nebula with 
the main features labelled. The cross  marks the location of the \oh
maser and the circle marks the field of view of the UNSWIRF aperture
(see Section~\ref{sec-unswirf}).}
\label{fig-radio}
\end{figure*}


\begin{figure*}
\centering
\includegraphics[height=8cm]{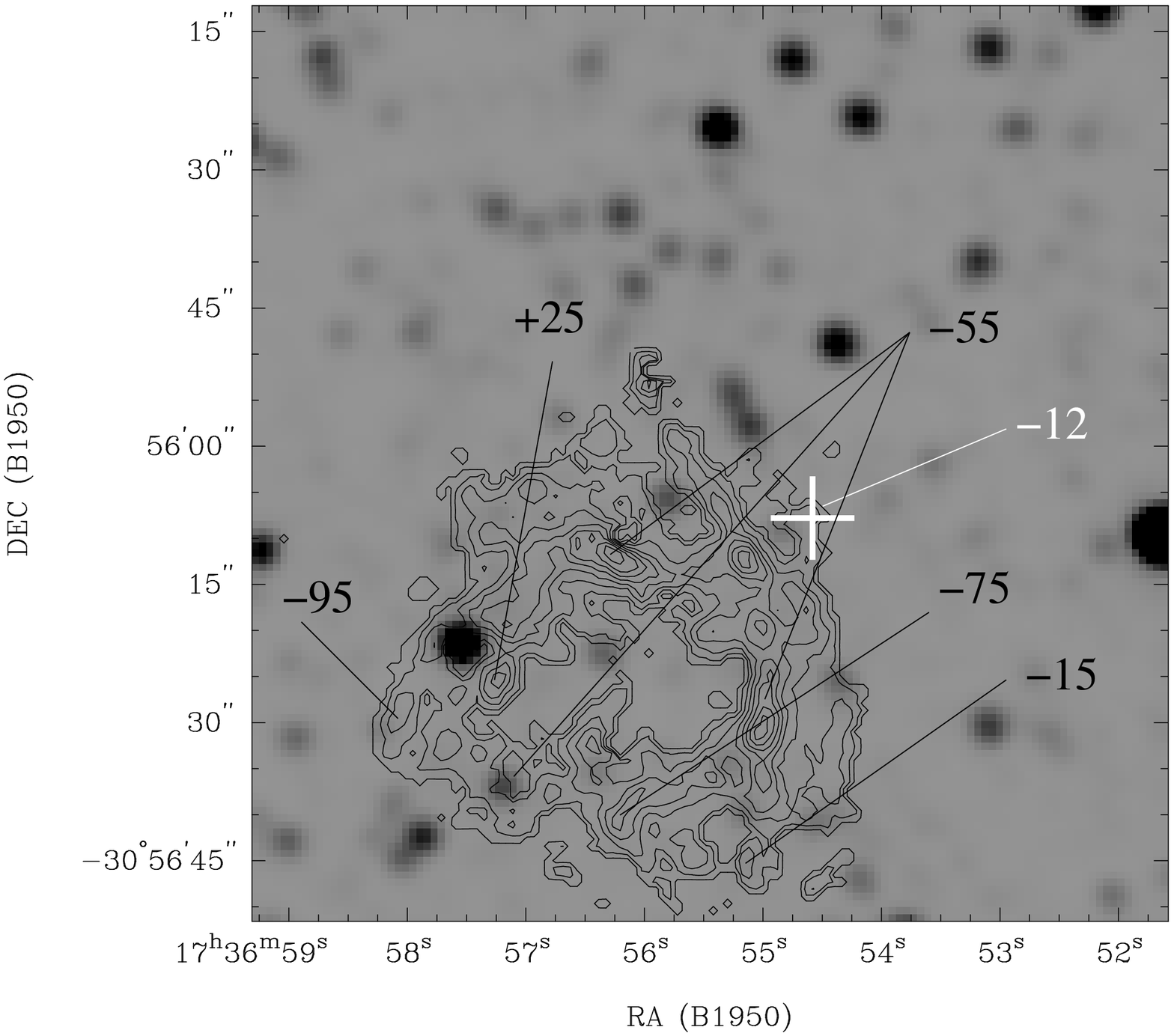}
\caption{Contours of the velocity-integrated \hone line
emission towards the \oh maser in the Tornado nebula, overlaid on the 
2MASS $K$-band greyscale image matching the size of the UNSWIRF field
of view. The contours are: 1.6, 4.7, 6.3, 7.9,
9.5, 11.1, 12.6, 14.2 \ee{-5}\ergs. The white cross marks the location
(and the velocity) of the \oh maser. Labelled are also the line centre 
velocities (in \kms)  of prominent features.}
\label{fig-h2}
\end{figure*}

\clearpage

\begin{figure*}
\centering
\includegraphics[height=8cm]{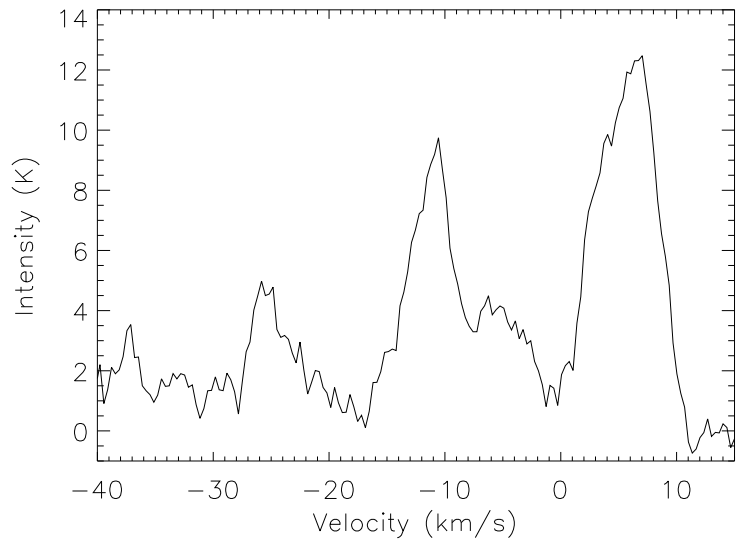}
\caption{\co\ 1--0 spectrum  observed 
towards the \oh maser in the Tornado nebula in the range $-40$ to
$+$15\kms\ demonstrates spectral features along the line of sight.}
\label{fig-spectra}
\end{figure*}


\begin{figure*}
\centering
\includegraphics[height=10cm]{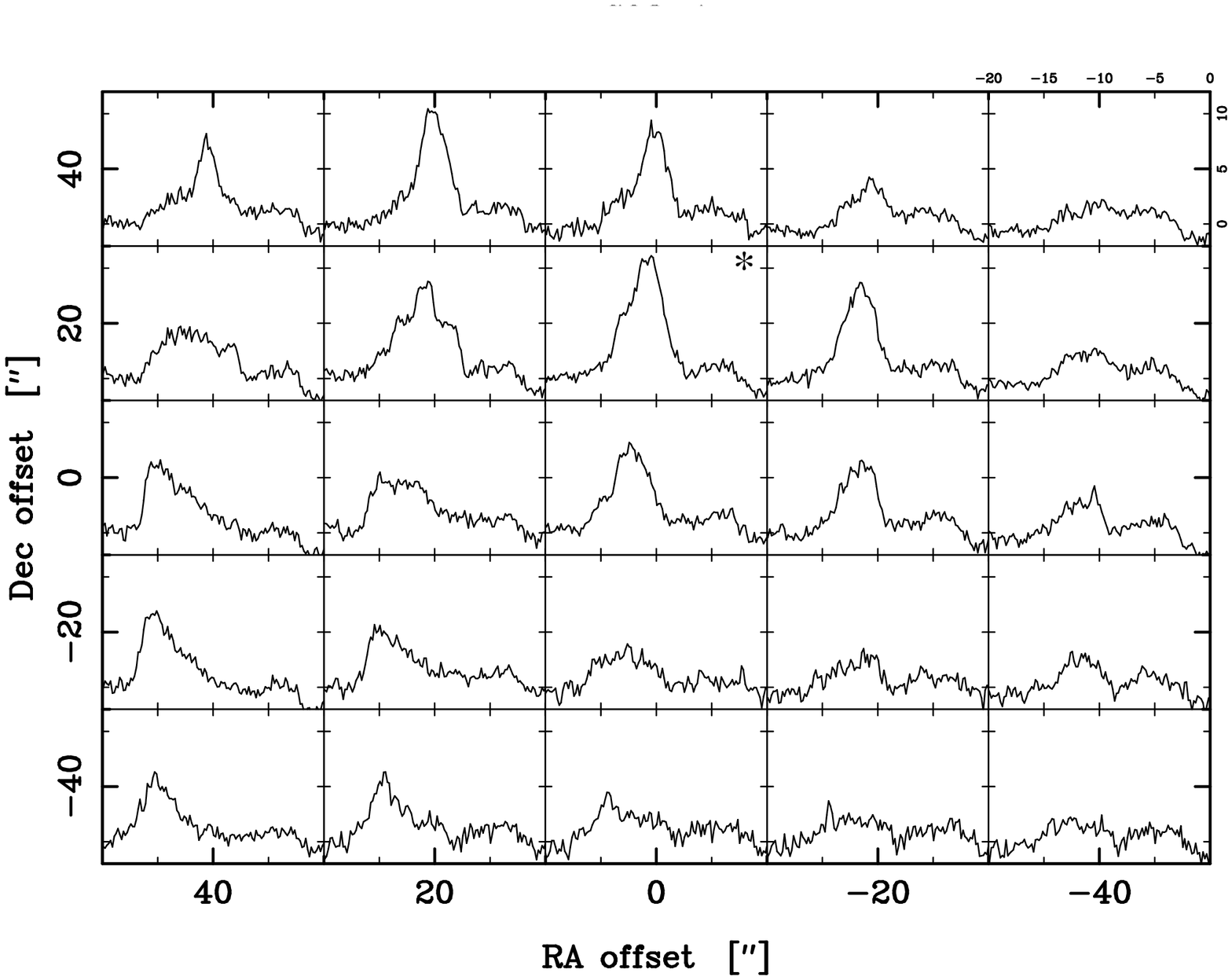}
\caption{\co\ 2--1 spectra taken over $2\arcmin \times 2\arcmin$ region
towards the \oh maser in the Tornado nebula. The (0,0) position is at 
${\rm RA}(1950)=17^{\rm h}~36^{\rm m}~54^{\rm s}$, 
${\rm Dec.}(1950)=-30\degr~56\arcmin~25\arcsec$, and the spectrum
closest to the \oh masers position is marked with a star. The scale is given
in the panel in the upper right; the velocity and temperature scales
are given in K and \kms, respectively.}
\label{fig-spectra-co}
\end{figure*}

\clearpage

\begin{figure*}
\centering
\includegraphics[height=10cm]{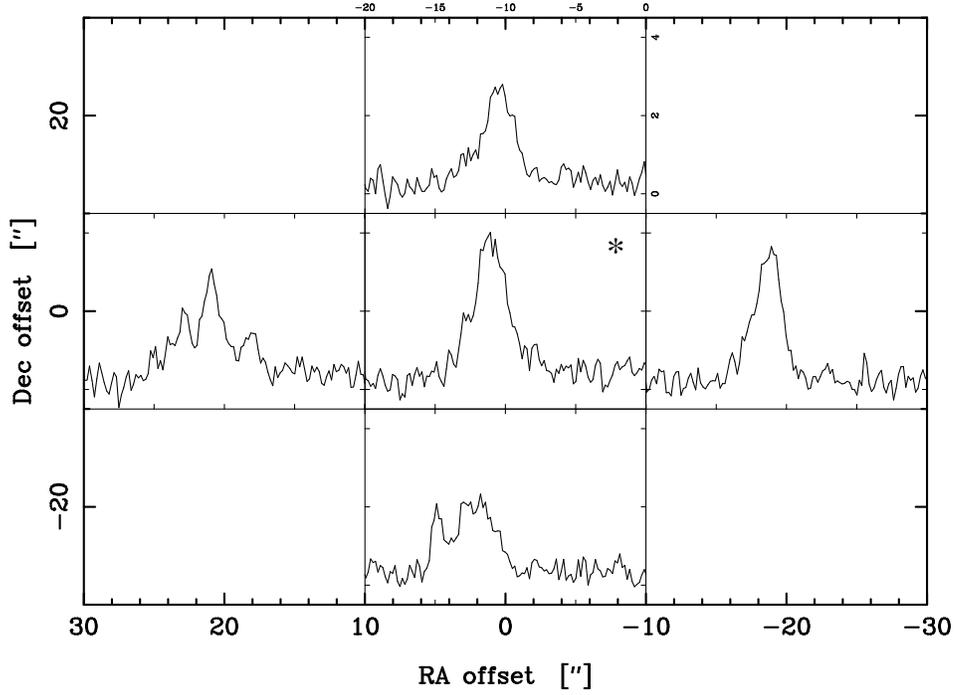}
\caption{\tco\ 2--1 spectra taken over $2\arcmin \times 2\arcmin$ region
towards the \oh maser in the Tornado nebula. The (0,0) position is at 
${\rm RA}(1950)=17^{\rm h}~36^{\rm m}~54^{\rm s}$, 
${\rm Dec.}(1950)=-30\degr~56\arcmin~05\arcsec$, and the spectrum
closest to the \oh masers position is marked with a star. The scale is given
in the upper panel; the velocity and temperature scales
are given in K and \kms, respectively.}
\label{fig-spectra-tco}
\end{figure*}


\begin{figure*}
\centering
\includegraphics[height=7cm]{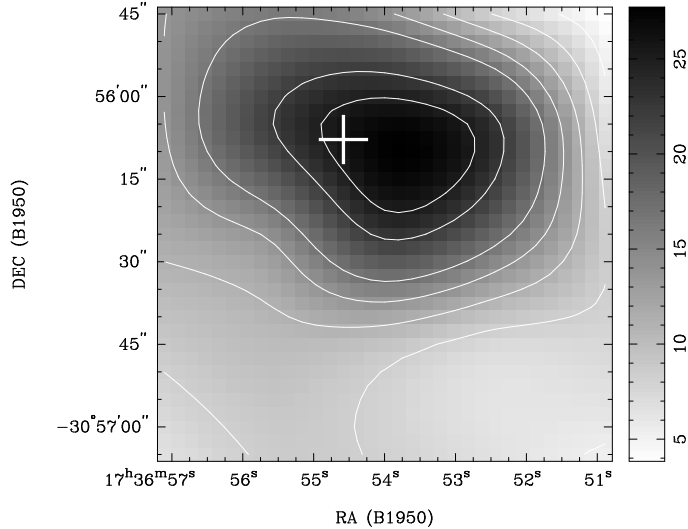}
\caption{Integrated \co\ 2--1 emission towards \oh maser in the
Tornado. The contours are: 8, 11, 14,
17, 22 and 25 K\kms. The cross marks the location of the \oh maser.}
\label{fig-co-map}
\end{figure*}

\clearpage

\begin{figure*}
\centering
\includegraphics[height=10cm]{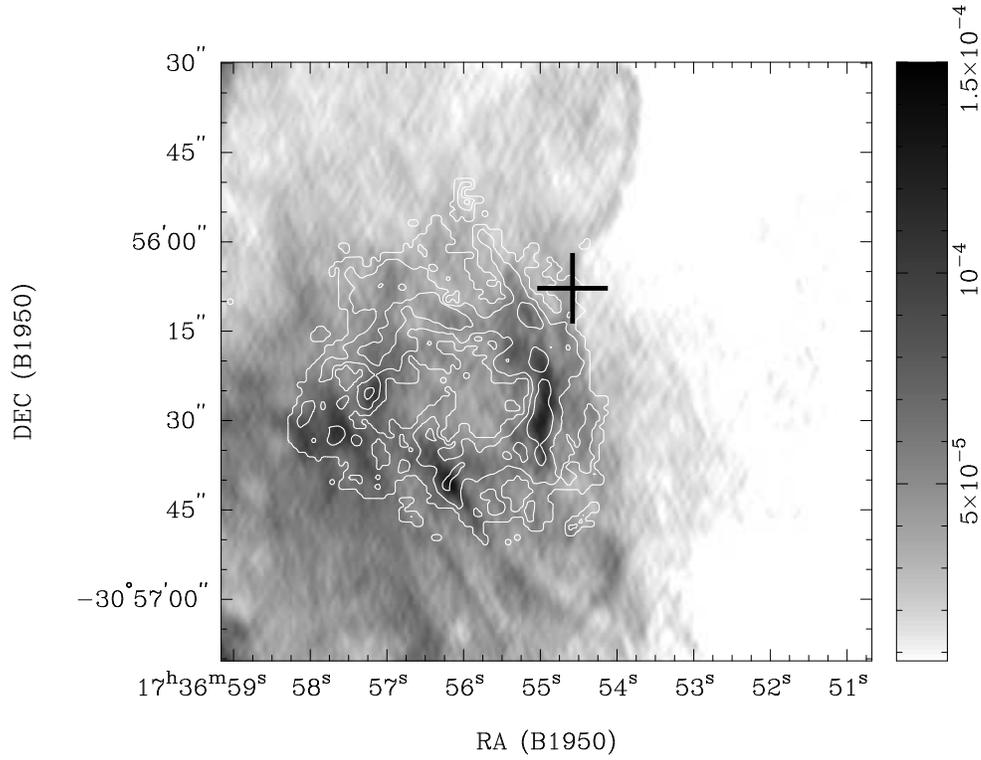}
\caption{Contours of \h\ 1--0 S(1) line emission superimposed on  
the 20~cm VLA grayscale image. The contour levels are 1.6, 6.3 and 
9.5\ee{-5}\ergs. The cross marks the location of the \oh maser.}
\label{fig-h2+radio}
\end{figure*}


\begin{figure*}
\centering
\includegraphics[height=6cm]{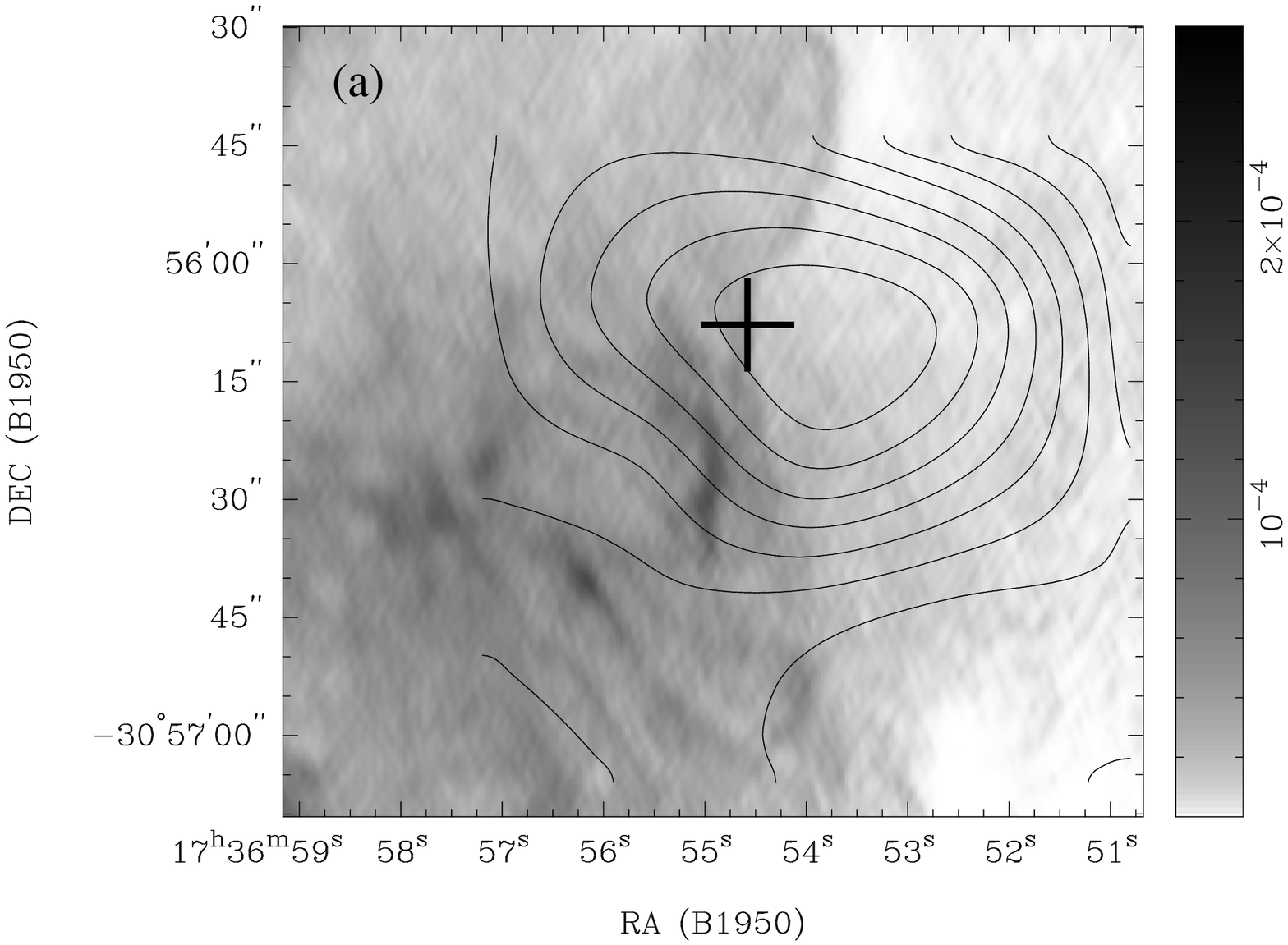}
\includegraphics[height=6cm]{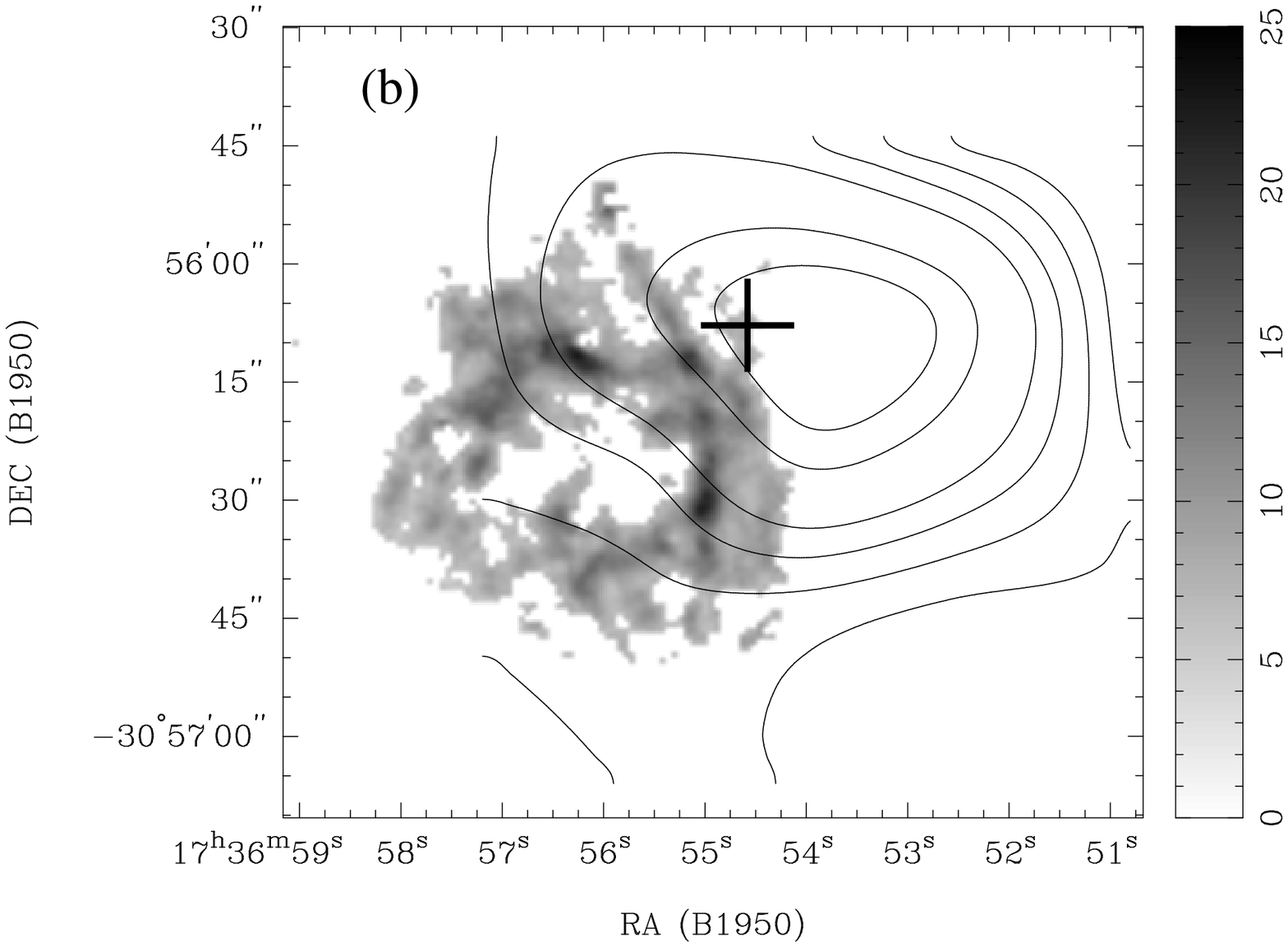}
\caption{Contours of the \co\ 2--1 line emission
 overlaid on ({\em a}) the 20~cm greyscale VLA image
and ({\em b}) the velocity-integrated \hone\ line emission
 in the Tornado nebula. The \co\ contours are the same as in
 Figure~\ref{fig-co-map}. 
The cross marks the location of the \oh maser.}
\label{fig-h2+co}
\end{figure*}


\end{document}